\documentclass[cits,pdftex]{PoS}
\usepackage{macros}
\usepackage{amssymb,amsmath}
\usepackage{mathrsfs}
\usepackage{booktabs,dcolumn}
\usepackage{grffile}         
\newcolumntype{C}{>{$}c<{$}} 
\title{%
Charm quark mass and D-meson decay constants from two-flavour lattice QCD
}

\author{%
\includegraphics[width=2.5cm]{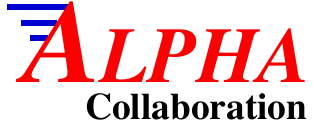}
\hfill\parbox{17.5mm}{\vspace{-1.0cm}\raggedleft\fns\it%
MS-TP-13-32\\
DESY 13-260
}
}

\ShortTitle{%
Charm quark mass and D-meson decay constants from $\nf=2$ lattice QCD
}

\author{\vspace{0.25cm}\speaker{%
Jochen Heitger}\\
Westf\"alische Wilhelms-Universit\"at M\"unster, 
Institut f\"ur Theoretische Physik,\\
Wilhelm-Klemm-Stra{\ss}e 9, D-48149 M\"unster, Germany\\
E-mail: \email{heitger@uni-muenster.de}
}
\author{%
Georg M. von Hippel\\
PRISMA Cluster of Excellence,\\
Johannes Gutenberg Universit\"at Mainz,
Institut f\"ur Kernphysik,\\
Johann-Joachim-Becher-Weg 45 , D-55099 Mainz, Germany\\
E-mail: \email{hippel@kph.uni-mainz.de}
}
\author{%
Stefan Schaefer and Francesco Virotta\\
NIC, DESY,
Platanenallee 6, D-15738 Zeuthen, Germany\\
E-mail: \email{stefan.schaefer@desy.de}
}

\abstract{%
\noindent
We present a computation of the charm quark's mass and the leptonic 
D-meson decay constants $f_{{\rm D}}$ and $f_{{\rm D}_{\rm s}}$ in 
two-flavour lattice QCD with non-perturbatively O($a$) improved Wilson 
quarks. 
Our analysis is based on the CLS configurations at two lattice spacings 
($a=0.065$ and $0.048$ fm, where the lattice scale is set by $f_{{\rm K}}$) 
and pion masses ranging down to $\sim 190$ MeV at $L\m_{\pi}\gtrsim 4$, 
in order to perform controlled continuum and chiral extrapolations with 
small systematic uncertainties.

\hfill\includegraphics[width=1.5cm]{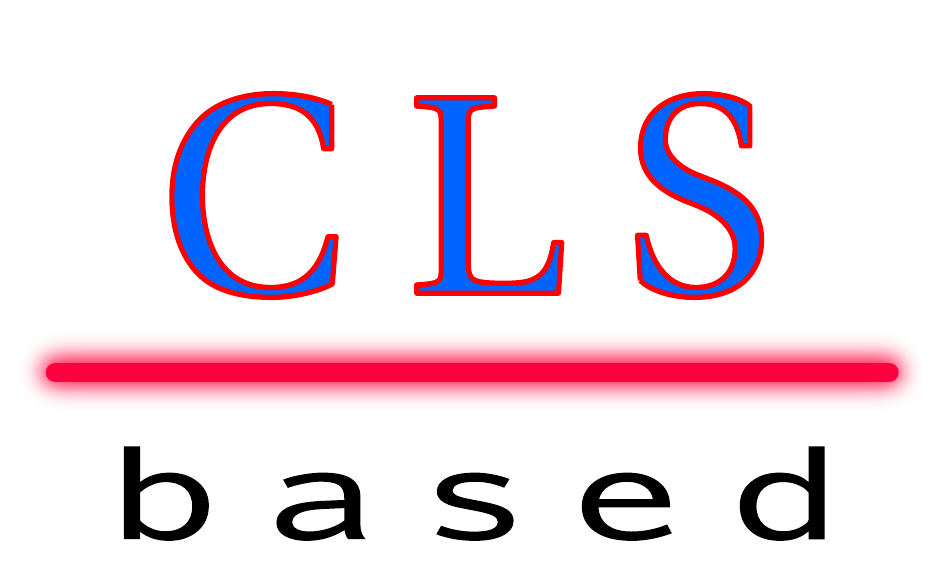}
}

\FullConference{%
31st International Symposium on Lattice Field Theory LATTICE 2013\\
July 29 - August 3, 2013\\
Mainz, Germany}
\begin{document}
%
%
\section{Introduction}
\label{Sec_intro}
\noindent
The decay constants $\fD$ and $\fDs$ entering the Standard Model expression 
for the leptonic decay widths of charged D-mesons are given by the
non-perturbative QCD matrix elements
\bea
\ketbra{0}{\overline{q}\gmu\gfv c}{\rmD_{q}(p)}=
{\rm i}{\fD}_q\,p_{\mu} \;,\quad
q={\rm d},{\rm s} \;.
\eea
Besides the importance of ${\fD}_q$ in extracting $|V_{{\rm c}q}|$ and
overconstraining the CKM matrix, there is continuing phenomenological 
interest in these decays, because a significant deviation between 
experimental and lattice results for ${\fD}_q$ could hint at New Physics in 
the flavour sector~\cite{Rosner:2012np}.
Extending our previous work~\cite{lat08:mcharmNf2} with an expanded set of 
ensembles with higher statistics, lighter sea quarks and largely reduced 
systematics (such as the uncertainty in the lattice scale), we here report 
results of a preliminary analysis of $f_{{\rm D}_{\rm (s)}}$ and of the charm 
quark mass.
\section{Computational setup and techniques}
\label{Sec_comp}
\noindent
Our measurements were performed on a subset of the Coordinated Lattice 
Simulations (CLS) gauge field ensembles, which are characterized by the 
Wilson plaquette gauge action and a sea of $\nf=2$ mass degenerate flavours 
of non-perturbatively $\Or(a)$ improved Wilson quarks.
The sea quark masses span a range corresponding to pion masses 
$(190\lesssim\mpi\lesssim 440)\,\MeV$, while the strange valence quark is
fixed to its physical value and we scan a range of charm valence quark 
masses around the physical charm quark mass.
Two values of the lattice spacing are considered: 
$a\in\{0.065\,\Fm\,,\,0.048\,\Fm\}$.
Suppression of finite-size effects is ensured by restricting ourselves to 
ensembles with $L\mpi\gtrsim 4$.
To generate the gauge configurations, either the
DD-HMC~\cite{ddhmc:luescher2,ddhmc:luescher3,deflat:luescher2,code:DDHMC} 
or the MP-HMC based on mass preconditioning~\cite{hmc:hasenb1,lat10:marina} 
was employed.
The parameters of the ensembles used are summarized in table~\ref{tab:cls}.
Statistical errors are estimated via a jackknife procedure, but will be 
double-checked in the final analysis by the method of~\cite{MCerr:ulli} 
studying autocorrelation functions.
%
\begin{table}[tb]
\begin{center}
\renewcommand{\arraystretch}{1.25}
\begin{tabular}{@{\extracolsep{0.2cm}}ccccccccc}
\toprule
id & $L/a$ & $\beta$ & $a\,[\,\Fm\,]$ & $\kapl$ & $\kaps$ & $m_{\pi}\,[\,\MeV\,]$ & \#\,cfgs \\
\midrule
E5 & $32$ & 5.3 & 0.065 & $0.136250$ & $0.135777(17)$ & 440 & $2\times 199$ \\
F6 & $48$ &     &       & $0.136350$ & $0.135741(17)$ & 310 & 250           \\
F7 & $48$ &     &       & $0.136380$ & $0.135730(17)$ & 270 & 547           \\
G8 & $64$ &     &       & $0.136417$ & $0.135705(17)$ & 190 & 820           \\
\hline
N5 & $48$ & 5.5 & 0.048 & $0.136600$ & $0.136262(08)$ & 440 & 238           \\
N6 & $48$ &     &       & $0.136670$ & $0.136250(08)$ & 340 & 1000          \\
O7 & $64$ &     &       & $0.136710$ & $0.136243(08)$ & 270 & 486           \\
\bottomrule
\end{tabular}
\caption{\sl%
Parameters of the ensembles used: 
ensemble label, spatial extent of the lattice in lattice units, bare 
coupling $\beta=6/g_0^2$, lattice spacing $a$, hopping parameters of the 
sea and (valence) strange quarks, the mass of the sea pion and the number 
of configurations employed.
All lattices have dimensions $T\times L^{3}$ with $T=2L$.
}\label{tab:cls}
\end{center}
\end{table}
%

The lattice spacings $a$, pion masses $\mpi$, pion decay constants $\fpi$
and values of the (quenched) strange quark's hopping parameter $\kaps$
are inferred from~\cite{scale:fK_Nf2}, where the scale is set through 
$f_{\rm K}$ at the ``physical point'' defined by 
$m_{\pi,{\rm phys}}=134.8\,\MeV$, $m_{{\rm K},{\rm phys}}=494.2\,\MeV$ and 
$f_{{\rm K},{\rm phys}}=155\,\MeV$ in the isospin-symmetric limit with QED 
effects removed.
For two ensembles (G8,N6), we had to perform a short linear interpolation
in $1/\kaps$ (at fixed charmed $\kapc$) to the physical value.

As for the hopping parameter of the (quenched) charm quark, $\kapc$, we fix
it by requiring the $\Ds$-meson mass to acquire its physical value,
$\mDs=m_{{\rm D_{\rm s}},{\rm phys}}=1968\,\MeV$, irrespective of the sea 
quark mass.
To this end, after choosing a few values in the vicinity of this target,
$(a\mDs)^2$ was interpolated linearly in $1/\kapc$ to 
$(am_{{\rm D_{\rm s}},{\rm phys}})^2$.
\section{Observables and analysis details}
\label{Sec_obs}
\noindent
For two mass non-degenerate valence quarks $r$ and $s$, we compute
correlators of the pseudoscalar density $P^{rs}=\psibar_r\gamma_5\psi_s$ and
the time component of the axial vector current 
$A_0^{rs}=\psibar_r\gamma_0\gamma_5\psi_s$ as
\bea
f_{\rm PP}^{rs}(x_0)=
-a^3\sum_{\vec{x}}\langle P^{rs}(x)P^{sr}(0)\rangle \;,\quad
f_{\rm AP}^{rs}(x_0)=
-a^3\sum_{\vec{x}}\langle A_0^{rs}(x)P^{sr}(0)\rangle \;.
\eea
These are evaluated using 10 $U(1)$ noise sources 
$\eta_t(x)=\delta_{t,x_0}\exp(i\phi(\vec x))$ located on randomly chosen time 
slices $t$~\cite{u1noise:rainer,u1noise:FM} so that solving the Dirac 
equation once for each noise vector 
$\zeta_t^r=Q^{-1}(m_{0,r})\eta_t= a^{-1}(D+m_{0,r})^{-1}\gamma_5\eta_t$ 
suffices to estimate the two-point functions projected onto zero momentum:
\bea
a^3 f_{\rm XP}^{rs}(x_0)=
{\T\sum_{\vec x}}\,\langle
[\zeta_t^r(x_0+t,\vec x)]^\dagger\Gamma\zeta_t^s(x_0+t,\vec x)
\rangle \;,\quad
\mbox{$\Gamma={\bf 1},\gamma_0$ for ${\rm X}={\rm P,A}$} \;,
\eea
where the average is over noise sources and gauge configurations.

\begin{figure}[tb]
\begin{center}
\vspace{-1.5cm}
\includegraphics[width=0.497\textwidth]{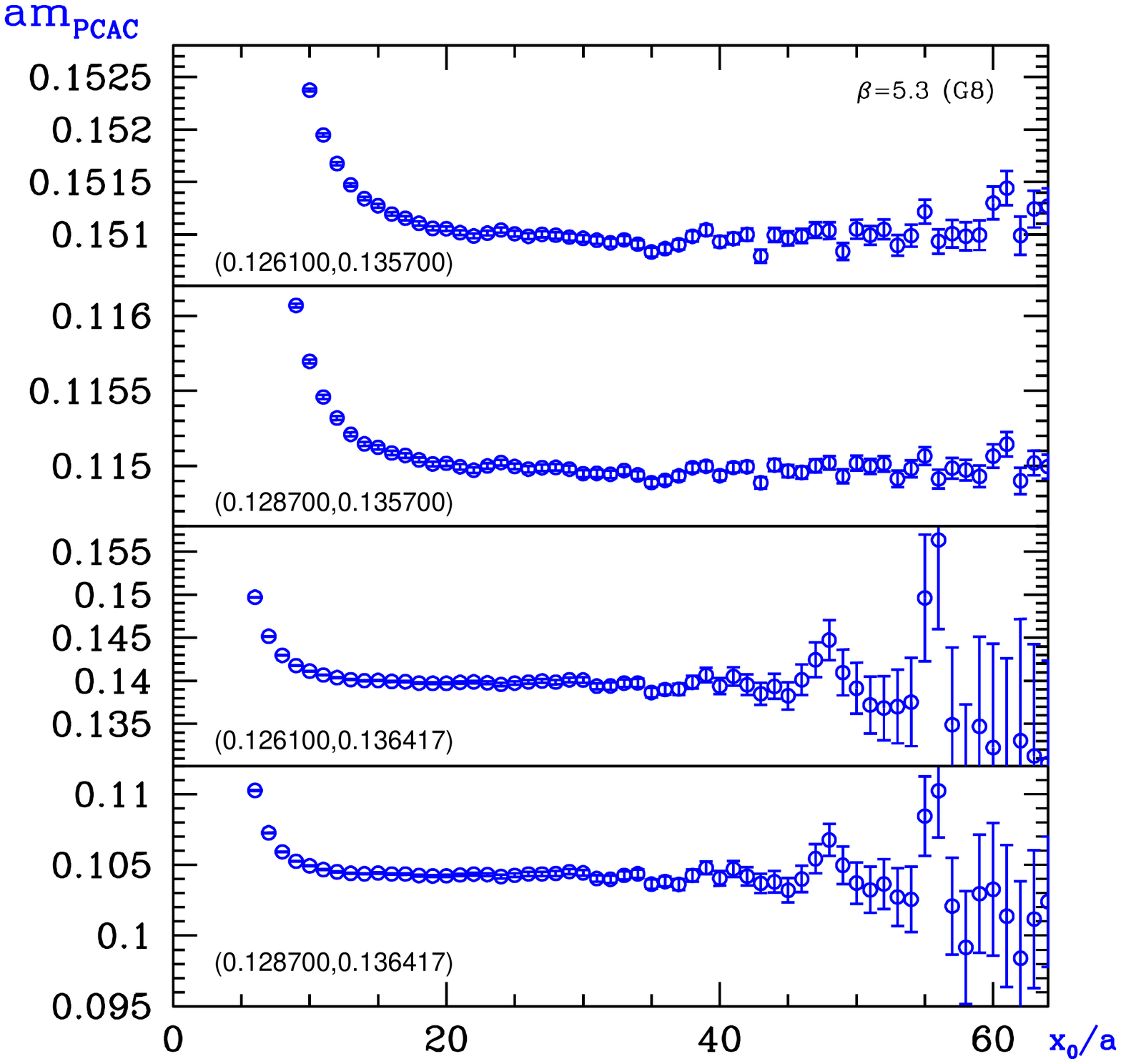}
\hfill
\includegraphics[width=0.497\textwidth]{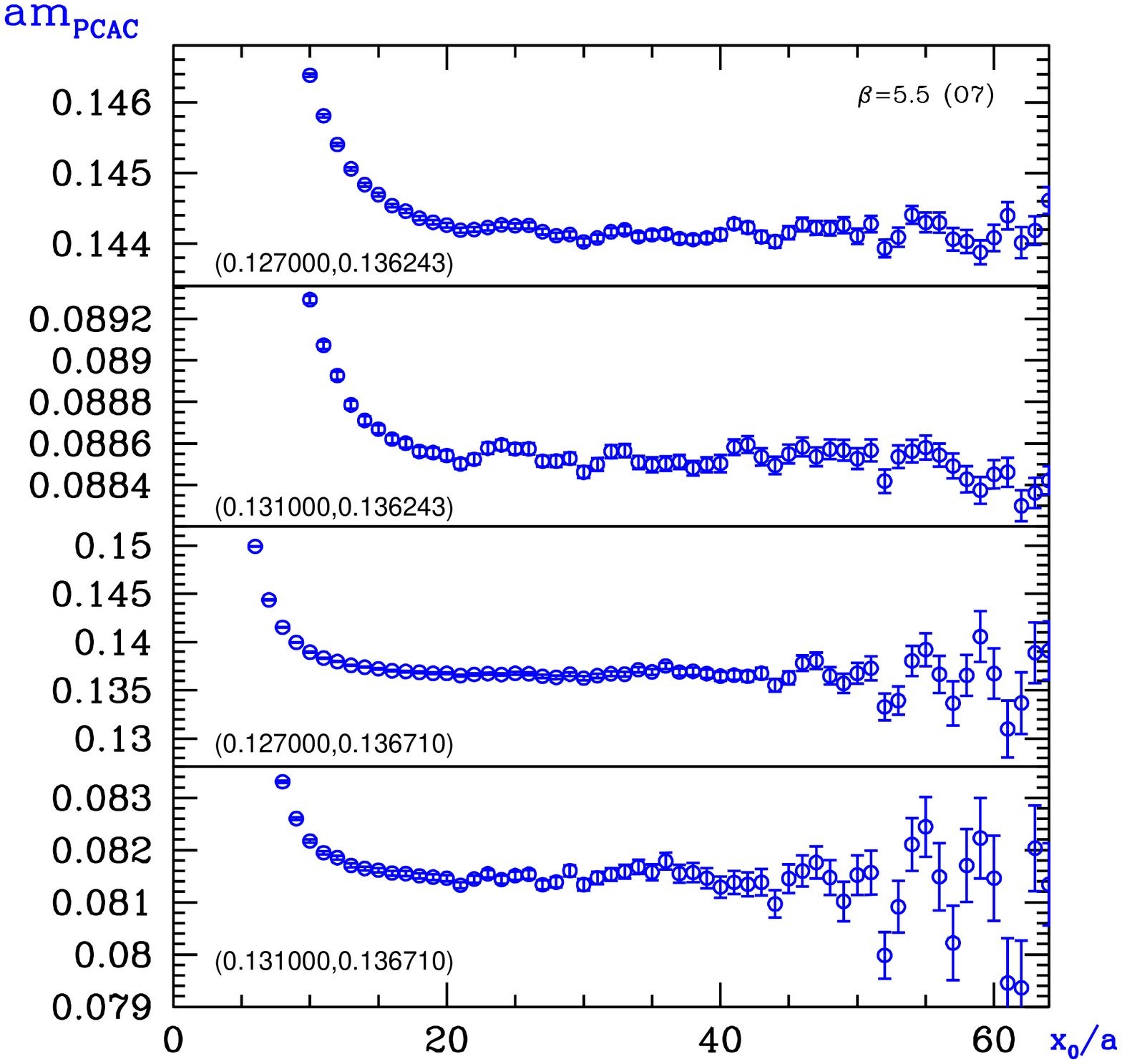}

\vspace{-2.0cm}
\caption{\sl%
Local PCAC quark masses for representative charm-strange (upper 2 panels) 
and charm-light (lower 2 panels) $\kappa$--combinations for the G8 and O7 
ensembles. 
Our bare quark mass estimates are obtained as averages over the plateau 
regions.
}\label{fig:mpcac}
\vspace{-0.5cm}
\end{center}
\end{figure}

The $\Or(a)$ improved effective average PCAC quark mass of flavours $r$ and 
$s$ is now defined as
\bea
\half\,(m_{rr}+m_{ss})(x_0)=m_{rs}(x_0)=
\frac{
\frac{1}{2}\,(\partial_0+\partial_0^*)f_{\rm AP}(x_0)
+\ca\,a\,\partial_0^*\partial_0 f_{\rm PP}(x_0)
}
{2f_{\rm PP}(x_0)} \;,
\label{mpcac}
\eea
where the improvement coefficient $\ca$ is non-perturbatively known 
from~\cite{impr:ca_nf2}.
For sufficiently large $x_0$, $m_{rs}(x_0)$ exhibits a plateau, over which we 
take a timeslice average to calculate $m_{rs}$.
Examples for two representative ensembles and various valence 
$\kappa$--combinations are shown in figure~\ref{fig:mpcac}.
%
%

The renormalized PCAC mass is then given by
\bea
\mr^{rs}=
\frac{\za(1+\babar a\msea+\batil a m_{rs})}
{\zp(1+\bpbar a\msea+\bptil a m_{rs})}\times m_{rs} \;,\quad
\mbox{$\msea=m_{{\rm l}{\rm l}}$ with ${\rm l}={\rm l(ight)}$} \;,
\label{mpcac_R}
\eea
where $\za$ and $\zp$ assume values from the non-perturbative determinations
in~\cite{scale:fK_Nf2,msbar:Nf2,impr:za_nf2_2,impr:babp_nf2}).
The $b$--coefficients, multiplying (very small) improvement terms, are known 
in 1-loop perturbation theory~\cite{scale:fK_Nf2,impr:pap5};
in particular, $\babar=\bpbar=0$ holds at this order.

\begin{figure}[tb]
\begin{center}
\vspace{-1.5cm}
\includegraphics[width=0.497\textwidth]{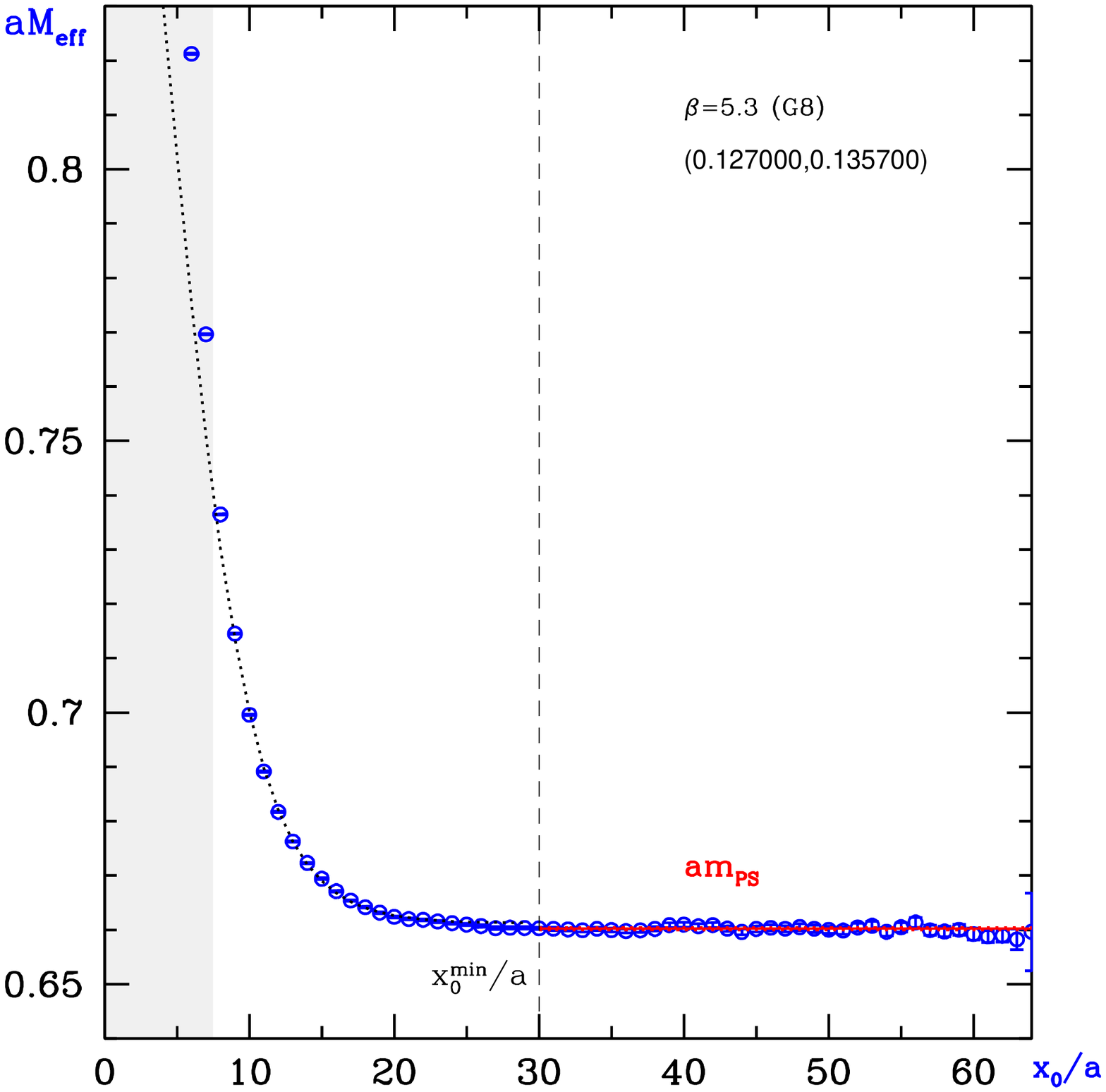}
\hfill
\includegraphics[width=0.497\textwidth]{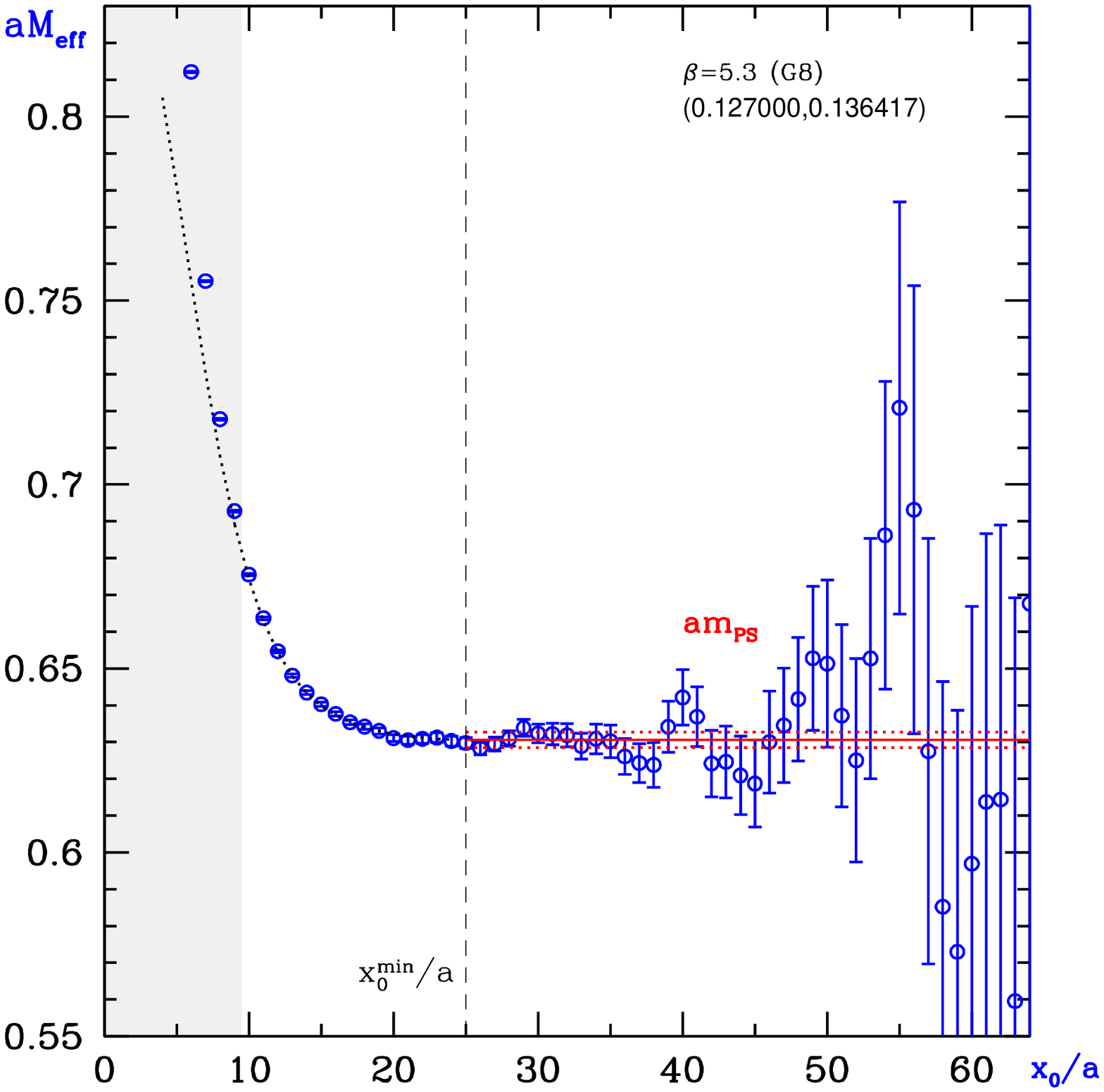}

\vspace{-2.0cm}
\caption{\sl%
The effective mass in the pseudoscalar channel extracted from $f_{\rm PP}$
for a representative charm-strange (left) and charm-light (right)
$\kappa$--combination for the G8 ensemble. 
A two-state fit to data outside the shaded area, where the fit function 
describes the data well given their accuracy, determines $x_0^{\rm min}$.
The result of the final one-state fit is given by the error band.
}\label{fig:mps}
\vspace{-0.5cm}
\end{center}
\end{figure}

Expressions for the pseudoscalar (PS) meson mass and its decay constant 
arise from the spectral decomposition for infinite $T$,
\bea
f_{\rm PP}(x_0)={\T\sum_{i=1}^\infty}\,c_i\,\Exp^{\,-E_i x_0} \;,\quad
E_1=\mps \;,\quad
\mbox{$E_{i\ge 2}$: excited states contributions} \;,
\eea
which decays exponentially for large time separations.
In this asymptotic regime, the decay constant is thus given by: 
\bea
\fps=
\za\,(1+\babar a\msea+\batil a m_{rs})\times\fps^{\rm bare} \;,\quad
\fps^{\rm bare}=
2\sqrt{2c_1}\,m_{rs}\,\mps^{-3/2} \;.
\label{fps}
\eea
Since in the actual analysis we face finite time extents $T$ and separations 
$x_0$, and hence particles running backwards in time and excited states,
we employ the following two-step procedure to fix the region 
$x_0\in[x_0^{\rm min},T-x_0^{\rm min}]$, in which the excited state 
contribution to $f_{\rm PP}$ can be neglected:
1.)~Determine $x_0^{\rm min}$ as the smallest $x_0$, where the excited state, 
estimated by a 2-state fit (including finite-$T$ effects) to
$f_{\rm PP}(x_0)=c_1\big[\Exp^{-E_1x_0}+\Exp^{-E_1(T-x_0)}\big]
+c_2\big[\Exp^{-E_2x_0}+\Exp^{-E_2(T-x_0)}\big]$, contributes less than
$1/4$ of the statistical uncertainty on the effective mass $M_{\rm eff}(x_0)$;
here the effective PS meson mass $M_{\rm eff}$ is defined as 
$\cosh\left[M_{\rm eff}(x_0-T/2)\right]
/\cosh\left[M_{\rm eff}(x_0+1-T/2)\right]=f_{\rm PP}(x_0)/f_{\rm PP}(x_0+1)$.
2.)~Perform a 1-state fit of the asymptotic exponential decay, restricted to 
this region, to extract $\mps=E_1$ and the leading coefficient $c_1$,
eventually entering the evaluation of the decay constants according to 
eq.~(\ref{fps}).
This is illustrated for our most chiral ensemble in figure~\ref{fig:mps}.
%
%
\section{Preliminary results}
\label{Sec_res}
\noindent
Unphysical pion masses and non-zero lattice spacings in our data are 
accounted for by employing joint chiral 
($m_{\pi}\to m_{\pi,{\rm phys}}=134.8\,\MeV$) and continuum limit ($a\to 0$) 
extrapolations.

\begin{figure}[tb]
\begin{center}
\vspace{-2.5cm}
\parbox{0.435\textwidth}{%
\includegraphics[width=0.485\textwidth]{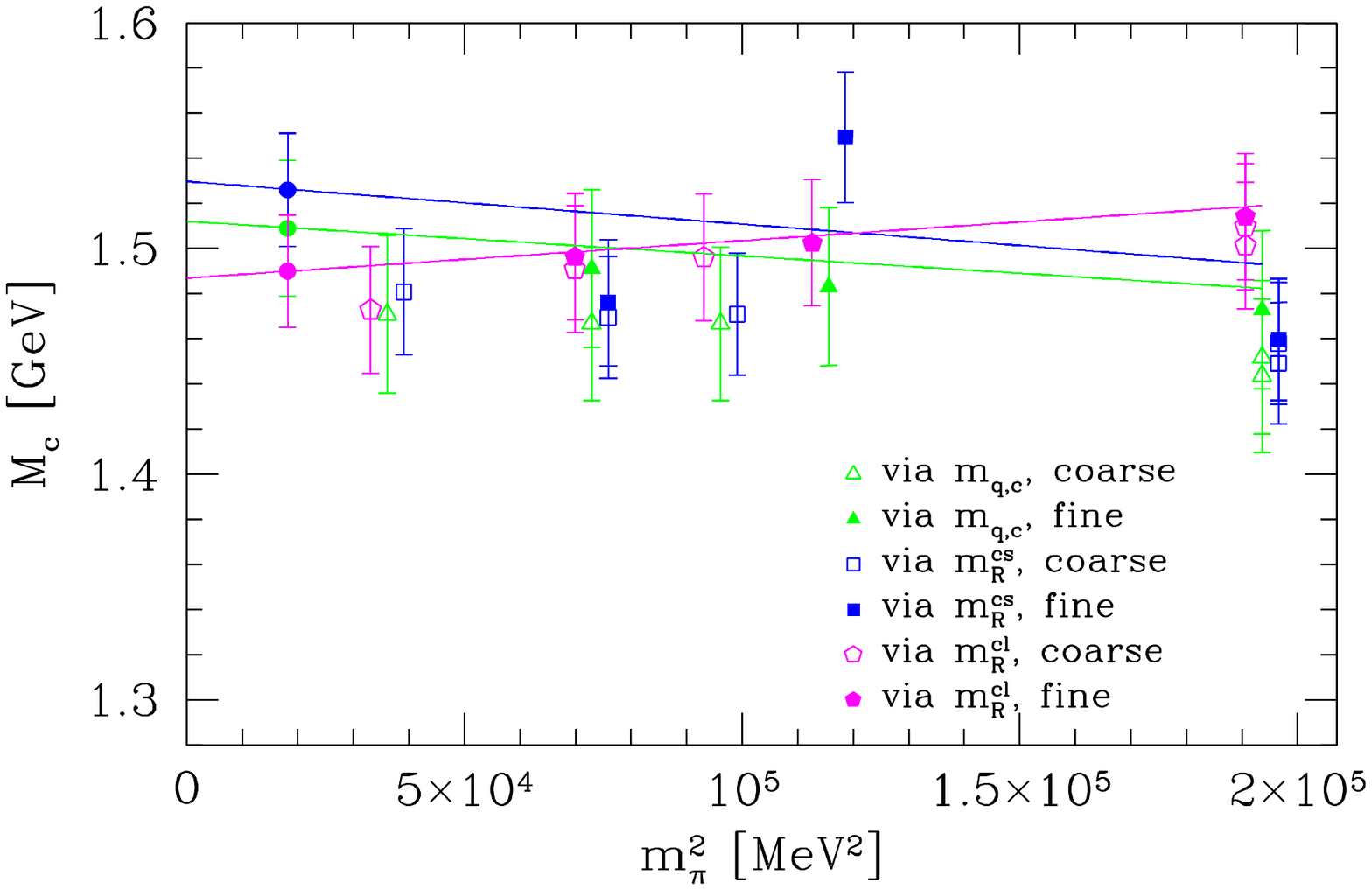}
}
\hfill
\parbox{0.555\textwidth}{%
\includegraphics[width=0.595\textwidth]{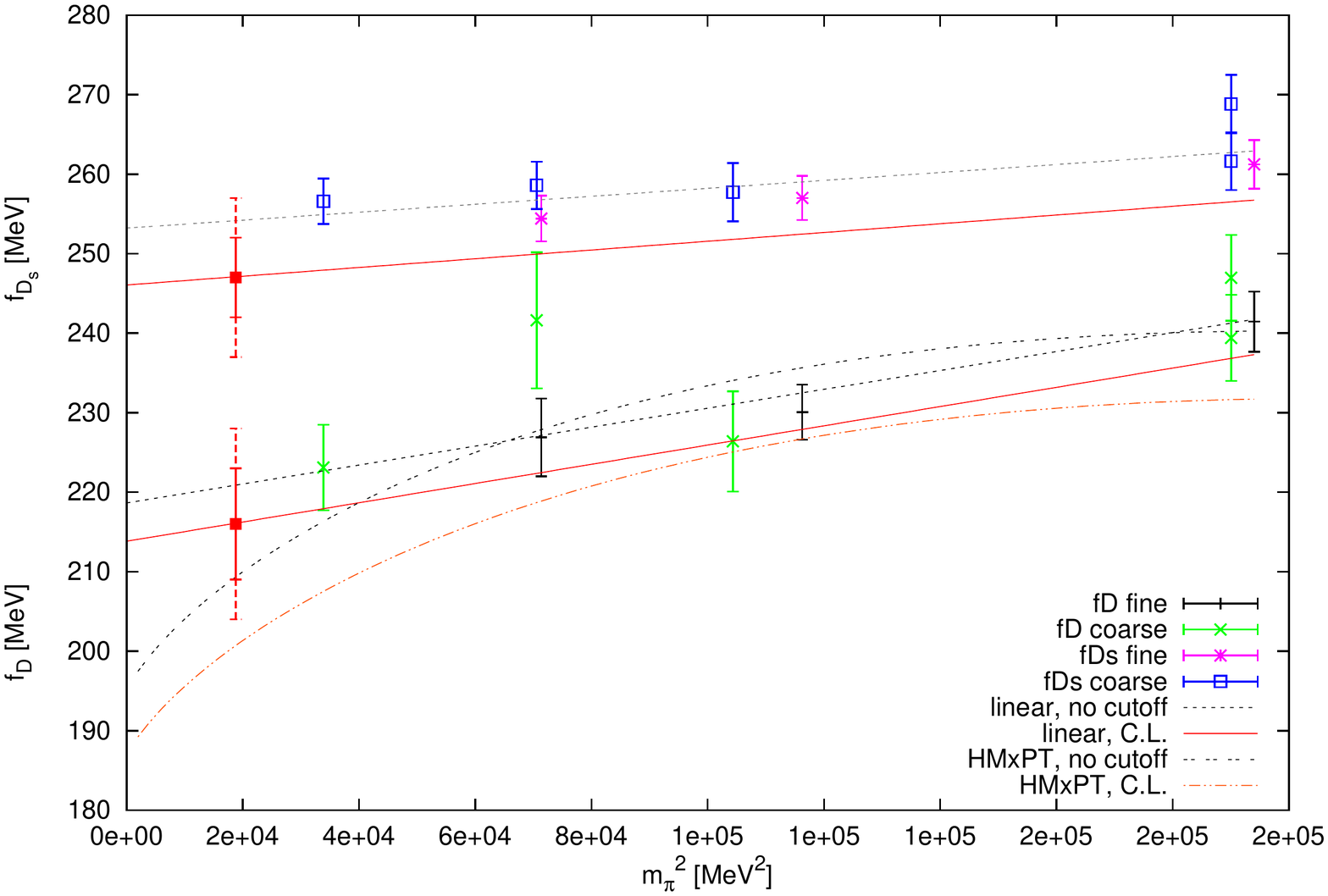}
}
\vspace{-2.75cm}
\caption{\sl%
\textit{Left:}
Joint chiral and continuum extrapolation of the RGI mass of the charm 
quark, using the definitions via the heavy-light PCAC relation 
(where $s={\rm s},{\rm l}$) and via the bare subtracted mass, 
eq.~(\protect\ref{mc}). 
\textit{Right:}
Joint chiral and continuum extrapolations to the physical point of $\fD$
and $\fDs$ to the fit ans\"atze in eqs.~(\protect\ref{fitansatzLO_fD+fDs}) 
and~(\protect\ref{fitansatzNLO_fD}).
Labels ``coarse'' and ``fine'' refer to $\beta=5.3$ and $5.5$, 
respectively, and for $\fDq{(s)}$ also fits neglecting cutoff effects are 
displayed. 
}\label{fig:mc+fdfds}
\vspace{-0.5cm}
\end{center}
\end{figure}

Assuming a linear dependence on the squared (sea) pion mass, our fit ansatz 
for the renormalized PCAC quark mass composed of a charm and a light valence
flavour ($s={\rm s},{\rm l}$) reads
\bea
\mr^{{\rm c}s}\left(\mpi,a\right)=
B+C\mpi^2+Da^2 \;.
\eea
In addition, we also consider the definition via the bare subtracted quark 
mass, $\mqc=m_{0,{\rm c}}-m_{\rm crit}$, so that we have three ways to obtain 
the renormalization group invariant (RGI) charm quark mass, 
$\Mc$~\cite{mcbar:RS02,fds:final}:
\bea
\half\,(\Mc+{M}_s)=
\frac{M}{\mbar}\,\mr^{{\rm c}s} \;;\quad
\Mc=
\frac{M}{\mbar}\,\frac{\za}{\zp}\,Z\left(1+\bm a\mqc\right)\mqc \;,\quad
\mqc=
\frac{1}{2}\Big(\frac{1}{\kapc}-\frac{1}{\kappa_{\rm crit}}\Big) \;.
\label{mc}
\eea
The universal factor $M/\mbar$, which translates the running mass at a given
scale to the RGI one, as well as the other renormalization and improvement 
factors entering here, are non-perturbatively known 
from~\cite{scale:fK_Nf2,msbar:Nf2,impr:za_nf2_2,impr:babp_nf2}.
The combined $\mpi^2$-- and $a^2$--dependence of the three definitions in 
eq.~(\ref{mc}) is shown in the left panel of figure~\ref{fig:mc+fdfds} to 
lead to consistent results in the joint chiral and continuum limit.
A more careful error analysis still to come, we consider the spread of these 
values as an upper limit for the overall uncertainty and quote as 
preliminary estimate for the charm quark's mass
\bea
\Mc=1.51(4)\,\GeV
\quad\Rightarrow\quad
\mcbMS\big(\mcbMS\big)=1.274(36)\,\GeV \;,
\eea
where $\Ms$ from~\cite{scale:fK_Nf2} and in the conversion to the $\MSbar$ 
scheme the known 4-loop anomalous dimensions of quark masses and 
coupling~\cite{Chetyrkin:1999pq,Melnikov:2000qh} together with $\lMSbar$ 
from~\cite{scale:fK_Nf2} were used.
%
%

For the D-meson decay constants, we adopt again fit ans\"atze linear 
in $\mpi^2$ (and $a^2$), while for $\fD$ we also model the sea quark 
dependence in a fit form inspired by partially quenched heavy meson chiral 
perturbation theory ($\hmcpt$)~\cite{hmchPT:G92,hmchPT:SZ96}, treating the 
charm quark as heavy, viz.
\bea
\fDq{(s)}\left(\mpi,a\right)
& = &  
b_{\rm (s)}+c_{\rm (s)}\,\mpi^2+d_{\rm (s)}\,a^2 \;,
\label{fitansatzLO_fD+fDs} \\
\fD\left(\mpi,a\right)
& = &
b'\left[1-\frac{3}{4}\,\frac{1+3\,\hat{g}^2}{(4\pi f_\pi)^2}
\,\mpi^2\ln\big(\mpi^2\big)\right]
+c\,'\mpi^2+d\,'\,a^2 \;;
\label{fitansatzNLO_fD}
\eea
$m_{\rm s}$--terms are assumed to be absorbed into the $\Or(\mpi^0)$ 
constants, because we work at fixed physical strange quark mass,
and $\widehat{g}=\gDstDpi=0.6$~\cite{DstarDpi:CLEO} is the
$\Dstar\rmD\pi$--coupling.
These combined chiral and continuum extrapolations to the physical point are 
depicted in the right panel of figure~\ref{fig:mc+fdfds}.
As can be seen from the fits, our data for $\fD$ are best described by
the linear extrapolation along eq.~(\ref{fitansatzLO_fD+fDs}), and we do not
see any evidence for the significance of the chiral logarithm-term in
eq.~(\ref{fitansatzNLO_fD}).
Therefore, we take the linear extrapolations as the central values to arrive
at our present results
\bea
\fDq{s}=
247(5)_{\rm stat}(5)_{\rm syst}\,\MeV \;,\quad
\fD=
216(7)_{\rm stat}(5)_{\rm syst}\,\MeV \;,\quad
\fDq{s}/\fD=
1.14(2)_{\rm stat}(3)_{\rm syst}
\label{res_fDs+fD}
\eea
and the difference to the $\hmcpt$ fit to account for a part of the 
systematic error of $\fD$.
Apart from the statistical errors, the quoted uncertainties also contain
a conservative estimate of the contribution from the scale setting.
\section{Conclusions and outlook}
\label{Sec_concl}
\noindent
The results for the charm quark mass and the D-meson decay constants from
our analysis are very well in line with computations of other groups, see, 
e.g., the recent summaries in~\cite{Aoki:2013ldr,lat13:bphys}. 
Note that by setting the scale through $f_{\rm K}$ we effectively compute
$\fDq{(s)}/f_{\rm K}$, where $\za$ in eq.~(\ref{fps}) (and thus also its 
error) drops out, but that it still re-enters indirectly by also fixing 
$\kaps,\kapc$ through $f_{\rm K}$~\cite{scale:fK_Nf2}.
This uncertainty, estimated conservatively so far, will likely decrease in
the final analysis.
%

\vspace{0.5cm}
{\footnotesize%
\noindent {\bf Acknowledgments.}
We thank Rainer Sommer for useful discussions and Nazario Tantalo for his
contribution at an early stage of this project.
This work is supported by the grant HE~4517/3-1 (J.~H.) of the Deutsche 
Forschungsgemeinschaft.
We are indebted to our colleagues in CLS for the joint production and use of 
the $\nf=2$ gauge configurations.
Most of our numerical simulations have been performed on the computers of 
the John von Neumann Institute for Computing at Forschungszentrum J\"ulich
(under project ID ``HCH09''), HLRN in Berlin and DESY, Zeuthen, and we thank 
these institutions for allocating computer time for this project and the 
computer center's staff for their technical support.
In particular, we also gratefully acknowledge the granted access to the HPC 
resources of the Gauss Center for Supercomputing at Forschungzentrum 
J\"ulich, Germany, made available within the Distributed European Computing 
Initiative by the PRACE-2IP, receiving funding from the European Community's 
Seventh Framework Programme (FP7/2007-2013) under grant agreement RI-283493.
}

%
\bibliography{lattice_ALPHA}
\bibliographystyle{JHEP-2_notitles}
%
%
\end{document}